\documentclass[]{article}
\usepackage[caption=false]{subfig}

\usepackage[natbibapa,nodoi]{apacite}
\usepackage{amsmath,amsfonts}

\newcommand{\fl}{{\beta, \eps}}
\newcommand{\fls}{{\beta,s}}
\renewcommand*{\d}{{\mathrm d}}

\newcommand{\Def}{\operatorname{Def}}

\newcommand{\Div}{\operatorname{div}}
\newcommand{\Lie}{\mathcal{L}}

\newcommand{\SDiff}{\mathcal{D}_\mu}
\newcommand{\argmin}{\operatorname*{arg\,min}}
\newcommand{\Ds}{\mathcal{D}}
\newcommand{\eps}{\varepsilon}

\newcommand{\Ric}{\operatorname{Ric}}
\newcommand{\rolap}{\tilde{\Delta}}

\newcommand{\vf}{V}

\newcommand{\vfd}{V_{\operatorname{div}}}
\newcommand{\R}{{\mathbb R}}

\title{Geodesic motion on the groups of diffeomorphisms with $H^1$ metric as geometric generalised Lagrangian mean theory}

\author{Marcel Oliver and Sergiy Vasylkevych}

\begin{document}

\maketitle
\abstract{%
Generalized Lagrangian mean theories are used to analyze the
interactions between mean flows and fluctuations, where the
decomposition is based on a Lagrangian description of the flow.  A
systematic geometric framework was recently developed by Gilbert and
Vanneste (J. Fluid Mech., 2018) who cast the decomposition in terms of
intrinsic operations on the group of volume preserving diffeomorphism
or on the full diffeomorphism group.  In this setting, the mean of an
ensemble of maps can be defined as the Riemannian center of mass on
either of these groups.  We apply this decomposition in the context of
Lagrangian averaging where equations of motion for the mean flow arise
via a variational principle from a mean Lagrangian, obtained from the
kinetic energy Lagrangian of ideal fluid flow via a small amplitude
expansion for the fluctuations.

We show that the Euler-$\alpha$ equations arise as Lagrangian averaged
Euler equations when using the $L^2$-geodesic mean on the volume
preserving diffeomorphism group of a manifold without boundaries,
imposing a ``Taylor hypothesis'', which states that first order
fluctuations are transported as a vector field by the mean flow, and
assuming that fluctuations are statistically isotropic.  Similarly,
the EPDiff equations arise as the Lagrangian averaged Burgers'
equations using the same argument on the full diffeomorphism group.
These results generalize an earlier observation by Oliver
(Proc. R. Soc. A, 2017) to manifolds in geometrically fully intrinsic
terms.}

\section{Introduction}

Averaging, in particular the description of the time evolution of
averaged quantities, is a perennial theme in fluid dynamics.  The
motivation derives from two initially disconnected themes: first, the
necessity to model turbulent flows in terms of Reynolds averaging or
large-eddy simulation; see, e.g., \cite{Alfonsi:2009:ReynoldsAN} and
\cite{Sag06} for surveys and detailed references.  And second, the
study of wave-mean flow interactions; see, e.g.,
\cite{Buehler:2014:WavesMF} and references therein.

While much of the theory and simulation of turbulence uses a
decomposition into mean and fluctuations (or coarse scale and fine
scale structure) in the Eulerian description of the flow, the
wave-mean flow community has looked at the problem from a Lagrangian
point of view for a long time.  In particular, \cite{andrews1978exact}
formulated a framework, the Generalized Lagrangian Mean (GLM), which
leads to nonlinear equations of motion for a suitably defined
Lagrangian mean of an ensemble of flows.  It has since become a
central ingredient for the theory of wave-mean flow interactions.

The idea of employing a Reynolds-type decomposition into mean flow and
turbulent fluctuations in the Lagrangian description of the fluid was
initially developed by \cite{holm1999fluctuation} and
\cite{MarsdenS:2001:GlobalWP,MarsdenS:2003:AnisotropicLA} who, under
certain closure assumptions, obtain the Euler-$\alpha$ (also known as
the Lagrangian averaged Euler) equations as the resulting mean flow
model.  \cite{soward2008derivation}, also see
\cite{RobertsS:2009:NavierSA}, obtain a similar, but not identical set
of equations using a different variational principle.

A recent paper by \cite{GilbertV:2016:GeometricGL} clarifies two
crucial aspects about Lagrangian mean theories.  First, such theories
can only fully consistent when they are written in geometrically
intrinsic terms; most crucially, the \cite{andrews1978exact}
generalized Lagrangian mean of a divergence free vector field is
generally not divergence free.  Thus, GLM theories should be
formulated intrinsically.  (We note that this has been done in the
work of Holm as well as Marsden and Shkoller, without spelling out the
general framework explicitly.)  Second, and most crucially, Gilbert
and Vanneste point out that once the notion of averaged map is
specified, for example as the Riemannian center of mass of an ensemble
of maps, the fluctuations of an ensemble of maps are fully determined
by an ensemble of vector fields; the maps can be reconstructed by
integration along geodesics on the group of maps.  This observation
let \cite{Oliver:2017:LagrangianAG} to reconsider the derivation of
the Euler-$\alpha$ equations and found that, for flows in Euclidean
space, it can be based on the following minimal set of assumptions:
\begin{enumerate}
\item[(a)] The averaged map is the minimizer of $L^2$-geodesic
distance,
\item[(b)] first order fluctuations are statistically isotropic, and
\item[(c)] first order fluctuations are transported by the mean flow
as a vector field.
\end{enumerate}
Hypothesis (c) was already used by
\cite{MarsdenS:2001:GlobalWP,MarsdenS:2003:AnisotropicLA} who refer to
it as the ``generalized Taylor hypothesis''.  The second order closure
stated by \cite{MarsdenS:2003:AnisotropicLA} is not assumed, but
arises as a necessary consequence of the geometric notion of averaged
map (a) together with (b) and (c).  Therefore, only the assumption of
isotropy of fluctuations (b) and the first order closure (c) are
modeling hypotheses which requires theoretical or empirical
verification.

In this paper, we show that these ideas extend to flows on manifolds
without boundaries and can be formulated in fully intrinsic terms.  We
also show that the same concept extends to the derivation of the
EPDiff equations as the Lagrangian averaged Burgers' equations.  The
significance of these results is twofold.  First, nontrivial manifolds
such as the sphere or spherical shells naturally arise in geophysical
fluid dynamics.  Second, it shows that the result of
\cite{Oliver:2017:LagrangianAG} is structurally robust and not tied to
special properties of Euclidean geometry.  Thus, for the first time,
we have achieved a fully intrinsic derivation of the Euler-$\alpha$
equations on non-Euclidean manifolds.

The crucial ingredient leading to a fully intrinsic derivation is the
correct interpretation of isotropy in the context of a non-flat
manifold.  It turns out that setting the fluctuation covariance tensor
to be a multiple of the inverse metric tensor results in all
curvature-induced terms in the average Lagrangian combine into the
Ricci Laplacian.

We make no claim about the validity of the Taylor hypothesis or the
usefulness of the Euler-$\alpha$ equations as a momentum closure for
turbulence.  A computational study of this question appears feasible,
even though it will not be easy and still requires a more careful
definition of the notion of ensemble mean than is necessary for the
purposes of this paper.  However, it is now clear that only the
dynamics of the first order fluctuation vector field would need to be
tracked.

The remainder of the paper is organized as follows.  In
Section~\ref{s.preliminaries}, we recall some basic notions from
differential geometry and the variational framework leading to the
Euler, the Euler-$\alpha$, Burgers', the EPDiff, and the Camassa--Holm
equations.  Section~\ref{s.geodesic} defines the geodesic mean of an
ensemble of maps.  In Section~\ref{s.averaging}, we explain the
concept of Lagrangian averaging, largely following the setup of
\cite{MarsdenS:2001:GlobalWP}.  The main closure assumption, the
generalized Taylor hypothesis, is introduced and applied to the
variational principle in Section~\ref{s.marsden-shkoller}.  The
following Section~\ref{s.isotropy} shows that this closure, under the
assumption of statistical isotropy and using the $L^2$-geodesic mean
on the full diffeomorphism group, implies the Euler-$\alpha$ or EPDiff
equations when considering, respectively, the group of volume
preserving diffeomorphisms or the full diffeomorphism group as
underlying configuration manifold.  For the Euler-$\alpha$ equations,
it is arguably more natural to use the geodesic mean with respect to
volume preserving geodesics, consistent with its underlying
configuration manifold.  This constraint introduces an additional
fictitious pressure term.  In Section~\ref{s.geodesic.vol}, we
demonstrate that this additional term does not contribute to the final
averaged Lagrangian.  For the sake of completeness, Section~\ref{s.ep}
recalls the derivation of the Euler-$\alpha$ and the EPDiff equations
as the Euler--Poincar\'e equations of the averaged Lagrangian.
Finally, in Section~\ref{s.boundaries}, we briefly discuss the
complications arising from boundaries.

\section{Notation and preliminaries}
\label{s.preliminaries}

Let $\Omega$ denote $n$-dimensional Euclidean space or a compact
$n$-dimensional Riemannian manifold without boundary, $g=g_{ij}$ be a
metric tensor on $\Omega$ with inverse $g^{ij}$, and
$\mu=\sqrt{g} \, \d x$ be the volume form on $\Omega$ induced by the
metric.

Let $\d$ and $\delta$ denote, respectively, the exterior derivative
and the co-differential associated to $g$.  We write $\nabla$ for the
Levi--Civita connection on $(\Omega,g)$ and $\nabla_v$ for the
covariant derivative in the direction of the vector field $v$.  Our
conventions for the Riemannian and Ricci curvature tensors,
correspondingly $R$ and $\Ric$, are
\begin{subequations}
\begin{gather}
  R(u,v)w
  \equiv \nabla_u \nabla_v w - \nabla_v \nabla_u w - \nabla_{[u,v]} w
  = R^i_{jkl} u^j v^k w^l \,, \\
  \Ric_{kl} = R^i_{ikl}
\end{gather}
\end{subequations}
for arbitrary vector fields $u$, $v$, and $w$ on $\Omega$, where
summation on repeated indices is implied in accordance with Einstein's
convention.

In the manifold context, it is necessary to distinguish between
different Laplace operators.  The \emph{rough Laplacian}
$\rolap = -\nabla^* \nabla$, where $\nabla^*$ is the $L^2$ adjoint of
$\nabla$, takes the form
\begin{equation}
  \label{e.l2-2}
  \rolap T = g^{ij} \, (\nabla_{e_i} \nabla_{e_j}
    - \nabla_{\nabla_{e_i}e_j}) \, T
\end{equation}
for an arbitrary tensor $T$.  The \emph{Hodge Laplacian} on vector fields
is given by
\begin{equation}
  \Delta u =[-(\d \delta + \delta \d ) u^\flat]^\sharp  
\end{equation}
where $\flat$ is a natural isomorphism between vector fields and
1-forms associate to $g$ and $\sharp=\flat^{-1}$. We recall that, by
the Weitzenb\"{o}ck formula
\citep{Petersen:2016:RiemannianG,Gay-BalmazR:2005:LiePS},
\begin{equation}
g(\Delta u,v) = g(\rolap u,v) - \Ric(u,v) \,. 
\end{equation}
Finally, we write $\Delta_R$ to denote the \emph{Ricci Laplacian},
\begin{equation} \label{e.riclap}
	g(\Delta_R u,v) = g(\rolap u, v) + \Ric (u,v) \,.
\end{equation} 
We remark than in Euclidean space, the differences between $\rolap$,
$\Delta$, and $\Delta_R$ vanish.

We write $\Ds(\Omega)$ to denote the group of $H^s$-class
diffeomorphisms of $\Omega$ and $\SDiff(\Omega)$ its volume preserving
subgroup.  For $s>n/2+1$, these groups are smooth infinite dimensional
manifolds in the $H^s$-topology
\citep{Palais:1968:FoundationsGN,EbinM:1970:GroupsDM} with tangent
spaces at the identity
\begin{subequations}
\begin{gather}
  \vf = \{ u \in H^s(\Omega, T \Omega) \colon u(x) \in T_x \Omega
           \text{ for } x \in \Omega \} \,, \\
  \vfd = \{  u \in \vf \colon \Div u = 0 \} \,.
\end{gather}
\end{subequations}
We write $\eta = \eta(t,x)$ to denote the flow of a time-dependent
vector field $u(t, \, \cdot \,) \in \vfd$, so that
\begin{equation}
  \label{e.flow}
  \partial_t \eta (t,x)=u(t,\eta(t,x)) 
\end{equation}
or $\dot \eta = u \circ \eta$ for short.  In this setting, the
equations of motions for many continuum theories can be viewed as
geodesic motion on one of the diffeomorphism groups with respect to a
particular choice of metric.  In other words, $u$ is a solution
whenever its associated flow $\eta$ is a stationary point of the
action
\begin{gather}
  S = \int_{t_1}^{t_2} L(\dot \eta, \eta) \, \d t
  \label{e.action}
\end{gather}	
with respect to variations $\delta \eta$ that are fixed at the
temporal end points.  In the context of this paper, we discuss the
following four cases.

As pointed out by \cite{Arnold:1966:GeometrieDG}, the \emph{Euler
equations} for the motion of an ideal incompressible fluid,
\begin{subequations}
  \label{e.euler}
\begin{gather}
  \dot u + \nabla_u u + \nabla p = 0 \,, \\
  \Div u = 0 \,,
\end{gather}
\end{subequations}
where $\nabla p \equiv \d p^{\sharp}$,  are the equations
for geodesic flow on $\SDiff$ with respect to the $L^2$-metric
\begin{equation}
  (u \circ \eta, v \circ \eta)_0
  = \int_{\Omega} g(u,v) \, \mu (x) \,.
\end{equation}
I.e., $u$ is a solution of \eqref{e.euler} whenever $\eta$ is a
stationary point of the action \eqref{e.action} with Lagrangian
\begin{gather}
  L(\dot \eta, \eta)
  = \tfrac{1}{2} \int_\Omega g(u,u) \, \mu (x)
  \equiv \tfrac{1}{2} \int_\Omega \lvert u \rvert^2 \, \mu (x) \,,
  \label{e.l2-lagrangian}
\end{gather}
where $u\subset \vfd $ and $\dot \eta \subset T \SDiff$ are related by
\eqref{e.flow}.

Similarly, \emph{Burgers' equations} 
\begin{equation}
  \label{e.burger}
  \dot u + \nabla_u u + (\nabla u)^T \cdot u +u \Div u = 0 \,,
\end{equation}
where the $(1,1)$-tensor $(\nabla u)^T$ is defined as the adjoint of
$\nabla u$ via
\begin{equation}
  g((\nabla u)^T \cdot v, w)\equiv g(\nabla_w u,v)
\end{equation}
for vector fields $u,v, w \in V$, is equivalent to the same
variational problem with Lagrangian \eqref{e.l2-lagrangian}, albeit
with configuration space $\Ds$ rather than $\SDiff$.

The \emph{Euler-$\alpha$ equations}
\begin{subequations}
  \label{e.eulera}
\begin{gather}
  \dot m + \nabla_u m + (\nabla u)^T \cdot m + \nabla p = 0 \,,\label{e.euleraa}\\
  m = u - \smash{\eps^2} \, \Delta_R u \,, \\ 
  \Div u = 0 
\end{gather}
\end{subequations}
are the equations for geodesic flow on the volume-preserving
diffeomorphism group $\SDiff$ with respect to a  right-invariant
$H^1$-metric. Their solutions are extremizers of the action $S$ upon
replacing the $L^2$-Lagrangian \eqref{e.l2-lagrangian} by
\begin{equation}
  L = \tfrac{1}{2} \int_\Omega |u|^2
      + 2 \, \eps^2 \, \lvert \Def u \rvert^2 \, \mu (x)\,,
  \label{e.LEulera}
\end{equation}
where $\Def u$ is the deformation tensor 
\begin{equation}
  \Def u =\tfrac{1}{2} (\nabla u+ \nabla u^T) 
\end{equation}
and $\lvert \Def u \rvert^2 = g(\Def u, \Def u)$ is defined by
extending metric $g$ to arbitrary $(1,1)$-tensors $S$ and $T$ via
\begin{equation}
  g(S,T) \equiv g_{ij} \, g^{kl} \, S^i_k \, T^j_l \,. 
\end{equation}

Finally, the \emph{EPDiff equations}
\begin{subequations}
  \label{e.epdiff}
\begin{gather} \label{e.epdiffa}
  \dot m + \nabla_u m + (\nabla u)^T \cdot m + m \Div u  = 0 \,,\\
  m = u - \eps^2 \, \Delta_R u \,, 
\end{gather}
\end{subequations}
describe geodesic flow on the full diffeomorphism group $\Ds$ with
respect to the right-invariant $H^1$-metric
\begin{equation}
  (u \circ \eta ,v \circ \eta)_1
  = \int_\Omega \lvert u \rvert^2
    + \eps^2 \, (\lvert \nabla u \rvert^2-\Ric(u,u)) \, \mu (x) \,.   
\end{equation}
Thus, solutions to \eqref{e.epdiff} are extremizers of the action $S$
corresponding to the Lagrangian
\begin{gather}
  L(\dot \eta, \eta)
  = \tfrac{1}{2} \int_\Omega \lvert u \rvert^2
    + \eps^2 \, (\lvert \nabla u \rvert^2-\Ric(u,u)) \, \mu (x)
  \label{e.Lepdiff}
\end{gather}	
on $\Ds$. For the sake of completeness, we sketch the derivation of
Euler-Poincar\'e equations \eqref{e.eulera} and \eqref{e.epdiff} from
their respective Lagrangians in Section~\ref{s.ep}. For missing
details, we refer the reader to \cite{HolmMR:1998:EulerPE},
\cite{Shkoller:2002:LagrangianAE}, and \cite{Gay-BalmazR:2005:LiePS}.

We note that Green's formulae for vector fields
$u,v \in \vf$, in the absence of boundaries, read
\begin{subequations}
\begin{gather}
  2 \int_\Omega g(\Def u, \Def v ) \, \mu (x)
  = -\int_\Omega g(\Delta_R u + \nabla \Div u,v) \, \mu (x)
  \label{e.g1}
\intertext{and}
  \int_\Omega g(\nabla u, \nabla v ) \, \mu (x)
  = - \int_\Omega g(\rolap u ,v) \, \mu (x)
\end{gather}
\end{subequations}
(see, for example, \citealt{Gay-BalmazR:2005:LiePS}).  Combining these
identities with \eqref{e.riclap}, we see that the Euler-$\alpha$
Lagrangian \eqref{e.LEulera} and the EPDiff Lagrangian
\eqref{e.Lepdiff} take the common form
\begin{equation}
  \label{e.LEuleraLap}
  L = \tfrac{1}{2} \int_{\Omega} g(u-\eps^2 \, \Delta_R,u) \, \mu (x) \,,
\end{equation}
the difference being that $u \in \vfd$ for the Euler-$\alpha$
equations and $u \in \vf$ for EPDiff.  In Section~\ref{s.ep}, we
sketch the derivation of the Euler-$\alpha$ and the EPDiff equations
from the Lagrangian in the form \eqref{e.LEuleraLap}.

On manifolds with boundaries, the two Lagrangians differ and the
expressions stated represent their most common form, for the
Euler-$\alpha$ equations, e.g., in \cite{MarsdenS:2001:GlobalWP}, and
for the EPDiff equations in \cite{HiraniMA:2001:AveragedTM} and
\cite{Gay-Balmaz:2009:WellPH}.

We finally remark that the EPDiff equations on $S^1$ or $\R$ reduce to
the peakon version of the Camassa--Holm equation (see, e.g.,
\citealt{CamassaH:1993:IntegrableSW}),
\begin{equation}
  u_t - \eps^2 \, u_{xxt}
  = -3 \, u \, u_x + 2 \, \eps^2 \, u_x \, u_{xx}
  + \eps^2 \, u \, u_{xxx} \,.
\end{equation}

\section{Geodesic mean}
\label{s.geodesic}
Let $\{\beta\}$ be an arbitrary index set, $\eps$ be a small parameter, and $u_\fl=u_\fl(x,t)$ denote the velocity field
corresponding to a single realization from an ensemble of flows on
$\Omega$.  It generates a flow
$\eta_\fl=\eta_\fl(x,t)$ via
\begin{equation}
\dot \eta_\fl = u_\fl \circ \eta_\fl 
\label{e.eta-eps} 
\end{equation}
with initial condition
$\eta_\fl \vert_{t=0} = \operatorname{id}$.  Now suppose that
the realizations can be decomposed into a averaged flow $\eta$ and a
fluctuating part $\xi_\fl$ via
\begin{equation}
\eta_\fl = \xi_\fl \circ \eta \,,
\label{e.decomposition}
\end{equation}
Both $\xi_\fl = \xi_\fl(x,t)$ and $\eta = \eta(x,t)$
are again time-dependent maps and we suppose that $\eta$ is generated
by a mean velocity field $u=u(x,t)$ via
\begin{equation}
\dot \eta = u \circ \eta 
\label{e.eta}
\end{equation}
where $\eta \vert_{t=0} = \operatorname{id}$.  When
$u_\fl \in \vf$, then $\eta_\fl \in \Ds$ and $\eta \in \Ds$.  When
$u_\fl \in \vfd$, then $\eta_\fl \in \SDiff$.  In this case, we seek
mean flows $\eta \in \SDiff$ that are also volume preserving. 

\cite{GilbertV:2016:GeometricGL} remark that flow maps $\eta_\fl$ are
points on the infinite dimensional group $\Ds(\Omega)$ or
$\SDiff(\Omega)$, hence it is possible to define the average map $\eta$
intrinsically, by utilizing the underlying geometric structure on the
group. They discuss several constructions for defining such averages.
In the following, we select those that remain fully within the
variational framework laid out in Section~\ref{s.preliminaries}: the
Riemannian center of mass of $\{\eta_\fl\}$ on $\Ds(\Omega)$ or
$\SDiff(\Omega)$. We recall the details of the construction below.

Suppose that we have a procedure $\langle \, \cdot\, \rangle$ for
averaging scalar quantities over the set $\beta$ which commutes with
spatial integration.  The
precise definition does not matter so long as the closure
assumptions, which we will introduce in the following sections, are
satisfied with respect to the induced notion of the mean.  Then, the
mean map $\eta$ on $\Ds(\Omega)$ is defined as the Fr\'echet mean
\begin{subequations}\label{e.mean-full}
	\begin{gather}
	\eta = \argmin_{\phi \in \Ds(\Omega)} \,
	\langle d^2_\eps(\phi,\eta_\fl) \rangle \,,
	\label{e.frechet-mean}
\end{gather}
where $d_\eps$ is a Riemannian distance function.  In principle, the
choice of metric is not unique.  However, we use the right-invariant
$L^2$-metric for the reason that it corresponds to the setting in
which the Euler equations and Burgers' equations, respectively,
describe geodesic flow.  Thus, the geodesic distance between two maps
$\phi, \psi \in \Ds(\Omega)$ is given by
\begin{gather}
	\label{e.metric-full}
	d_\eps^2(\phi,\psi) 
	= \inf\limits_{\substack{\gamma_s \colon [0,\eps] \rightarrow \Ds
			\\ \gamma_0=\phi \,, \gamma_1=\psi}} \int_0^\eps \int_{\Omega}
	g(\gamma_s', \gamma_s') \, \mu (x) \, \d s \,.
	\end{gather}
\end{subequations}
Here and in the following, the prime symbol denotes a derivative with
respect to $s$, which we think of as an arclength-like parameter.
Thus, the scaling introduced into \eqref{e.metric-full} indicates that
we will consider small fluctuations lying on a sphere of Riemannian
radius $O(\eps)$ about the mean.  \cite{GilbertV:2016:GeometricGL}
show that a single realizations $\eta_\fl$ is reached from $\eta$ by
integrating the transport equation
\begin{subequations}
	\label{e.transport} 
	\begin{gather}
	\label{e.transport.a}
	w_\fls' + \nabla_{w_\fls} w_\fls = 0 \,, 	
\end{gather}
in fictitious time $s$ from $s=0$ to $s=\eps$, together with a
constraint on the initial condition,
\begin{gather}
  \langle w_\fls \rangle \big|_{s=0} = 0 \,.
  \label{e.transport.b}
\end{gather}
\end{subequations}
The geodesic $\eta_\fls$ connecting $\eta$ and $\eta_\fl$ then is the
curve in $\Ds (\Omega)$ satisfying
\begin{equation}
\eta_\fls' = w_\fls \circ \eta_\fls 
\label{e.ws}
\end{equation}
with the initial condition $\eta_\fls \vert_{s=0} = \eta$. 

When the configuration space is the volumorphism 
group $\SDiff(\Omega) \subset \Ds(\Omega)$, there are two options to
define the mean.  We can either use the Fr\'echet mean with the
Riemannian distance inherited from $\Ds$, 
\begin{gather}
\eta = \argmin_{\phi \in \SDiff(\Omega)} \,
\langle d^2_\eps(\phi,\eta_\fl) \rangle \,,
\label{e.frechet-mean-vol-full}
\end{gather} 
or use Riemannian distance intrinsic to $\SDiff$, so that
\begin{subequations} \label{e.mean-vol}
\begin{gather}
  \eta = \argmin_{\phi \in \SDiff(\Omega)} \,
  \langle d_{\eps,\mu}^2(\phi,\eta_\fl) \rangle 
  \label{e.frechet-mean-vol} 
\end{gather}
with
\begin{gather}
  \label{e.metric-vol}
  d_{\eps,\mu}^2(\phi,\psi) 
  = \inf\limits_{\substack{\gamma_s \colon [0,\eps] \rightarrow \SDiff
      \\ \gamma_0=\phi \,, \gamma_1=\psi}} \int_0^\eps \int_{\Omega}
    g(\gamma_s', \gamma_s') \, \mu (x) \, \d t \,.
\end{gather}
\end{subequations}
In the first case, the fluctuation vector fields $w_\fls$ satisfy the
same transport equation \eqref{e.transport.a} together with the
constraint on the initial condition 
\begin{equation} \label{e.opttransic}
\left \langle w_\fls \right \rangle|_{s=0} = \nabla \psi 
\end{equation}
for some function $\psi$. In the
second case, the fluctuation vector fields satisfy an incompressible
Euler equation in fictitious time $s$,
\begin{subequations}
  \label{e.transportvol}
\begin{gather} 
  w_\fls' + \nabla_{w_\fls} w_\fls + \nabla \phi_\fls = 0 \,, 
  \label{e.transportvol.a} \\
  \Div w_\fls = 0 \,,
\end{gather}
\end{subequations}
with initial conditions constrained by \eqref{e.transport.b}.
Surprisingly, as we shall demonstrate, both choices lead to the same
averaged Lagrangian.

\section{Lagrangian averaging of geodesic flows on diffeomorphism groups}
\label{s.averaging}

The advantage of using Riemannian center of mass as the definition of
mean flow is that the averaged equations inherit material conservation
laws from the underlying system.  \cite{GilbertV:2016:GeometricGL}
derive averaged equations of motion by using the map $\xi_\beta$ to
pull back the momentum one-form to the the mean flow, then applying
averaging.  The resulting equations still need modeling in the form of
a relation between the averaged momentum one-form and the mean
velocity.  In this paper, we take a different approach: we average the
underlying system Lagrangian over the set of fluctuations to some
order in small fluctuation expansion first and compute the resulting
Euler--Poincar\'e equations from the resulting averaged Lagrangian
second.  This approach has been pioneered by
\cite{holm1999fluctuation} and
\cite{MarsdenS:2001:GlobalWP,MarsdenS:2003:AnisotropicLA} without
reference to the concept of geodesic mean.  Our approach differs from
the earlier works in that we average the Lagrangian over an ensemble
of fluctuations around the Riemannian center of mass while Marsden and
Shkoller average over a ball around a point. The approaches would be
equivalent if the center of Riemannian sphere were always its
Riemannian center of mass which, depending on the choice of measure on
the sphere, is generally not the case.

We proceed perturbatively, with the amplitude of fluctuations $\eps$
as small parameter.  It is convenient to work in the Eulerian
representation.  Let $L_\eps \equiv L(\eta_\fl, \dot \eta_\fl)$ denote
the $L^2$-Lagrangian for the Euler equations or Burgers' equations for
a \emph{single realization} of the flow, defined, respectively, on
$\SDiff$ or $\Ds$.  We treat both cases in parallel, pointing out
important differences along the way.  We recall the underlying kinetic
energy Lagrangian,
\begin{equation}
  \label{e.l-eps}
  L_\eps = \tfrac12 \int_\Omega g(u_\fl, u_\fl) \, \mu (x) \,,
\end{equation}
and expand $u$ in powers of $\varepsilon$, writing
\begin{equation}
u_\fl
= u + \varepsilon \, u_\beta'
+ \tfrac12 \, \varepsilon^2 \, u_\beta'' + O(\varepsilon^3) \,.
\label{e.u-expansion}
\end{equation}
Note that, to simplify notation, we read the absence of the index
$\eps$ as evaluation at $\eps=0$ so that, in particular,
$w_\beta \equiv w_\fls \vert_{s=0} = w_\fl \vert_{\eps=0}$.
Then,
\begin{align}
L_\varepsilon
& = \tfrac12 \int_\Omega
\bigl[
\lvert u \rvert^2
+ 2 \, \varepsilon \, g( u, u_\beta')
+ \varepsilon^2 \,
(\lvert u' \rvert^2 + g(u, u_\beta'')
\bigr] \, \mu (x)
+ O (\varepsilon^3) 
\notag \\
& \equiv L_0 + \varepsilon \, L_1 + \tfrac12 \, \varepsilon^2 \, L_2
+ O (\varepsilon^3) \,.
\label{e.Leps}
\end{align}
Truncating terms at $O(\eps^2)$ and taking the average, we introduce
an \emph{averaged Lagrangian} $\bar L$,
\begin{align}
\bar L & \equiv \tfrac12 \,
\biggl\langle
\int_\Omega
\lvert u \rvert^2
+ 2 \, \varepsilon \, u \cdot u_\beta'
+ \varepsilon^2 \,
\bigl(\lvert u_\beta' \rvert^2 + u \cdot u_\beta'' \bigr) \, \mu (x)
\biggr\rangle 
\notag \\
& = \tfrac12 \int_\Omega
\lvert u \rvert^2
+ 2 \, \varepsilon g(\, u, \langle u_\beta' \rangle)
+ \varepsilon^2 \,
\bigl(
\langle \lvert u_\beta' \rvert^2 \rangle
+ g(u, \langle u_\beta'' \rangle)
\bigr) \, \mu (x) \,.
\label{e.L}
\end{align}
This form of the averaged Lagrangian needs closure, i.e., we need to
express the averaged quantities in terms of mean quantities.  To so
so, we first note that $\eps$-derivatives of $u_\fl$ are not
independent of the perturbation vector fields $w_\fl$.  Indeed, recall
that
\begin{equation}
  \dot \eta_\fl = u_\fl \circ \eta_\fl 
  \label{e.us} 
\end{equation}
with the initial condition $\eta_\fl \vert_{t=0} = \operatorname{id}$.
Differentiating \eqref{e.ws} with respect to $t$, \eqref{e.us} with
respect to $s$ and equating the resulting mixed partial derivatives,
we obtain
\begin{gather}\label{e.u-prime-eps}
u_\fls'
=      \dot w_\fls + \nabla_{u_\fls} w_\beta \, 
- \nabla_{w_\fls} u_\fls \, 
\equiv \dot w_\fls + \Lie_{u_\fls} w_\fls \,,
\end{gather}
where we write $\Lie_{u}w$ to denote the Lie derivative of the vector
field $w$ in the direction of $u$.  Differentiating \eqref{e.u-prime-eps} and
evaluating at $\varepsilon=0$, we obtain the following
expressions for the coefficients of the $u_\fl$-expansion in
terms of the fluctuation vector fields $w_\beta$:
\begin{subequations}
	\begin{gather}
	\label{e.uprime.a}
	u_\beta'
	=      \dot w_\beta + \Lie_u w_\beta \,,  \\
	u_\beta'' = \dot w_\beta' + \Lie_u w_\beta' + \Lie_{u'} w_\beta \,.
	\end{gather}
	\label{e.uprime}
\end{subequations}
These relations show that once a notion of mean map is imposed,
represented by \eqref{e.mean-full}, \eqref{e.frechet-mean-vol-full},
or \eqref{e.mean-vol}, the problem remains in need of a single closure
condition: we are still free to choose an evolution equation for the
\emph{first order} fluctuation vector field $w_\beta$.  This will be
discussed in the next section.

\section{Generalized Taylor hypothesis}
\label{s.marsden-shkoller}
We choose a closure condition in the form 
\begin{equation}
\dot w_\beta + \Lie_{u} w_\beta = 0 \,.
\label{e.first-order-taylor}
\end{equation}
The expressions for the first and second order fluctuations of the
velocity field \eqref{e.uprime} then reduce to
\begin{subequations} \label{e.up-upp}
\begin{gather}
\label{e.up}  u_\beta' = 0 \,, \\ 
  u_\beta'' = \dot w_\beta' + \Lie_{u} w' \,. 
  \label{e.upp-reduced}
\end{gather}
\end{subequations}

Up until this point the procedure for Euler and Burgers' equation was
completely identical and it did not matter whether the map averaging
is defined by \eqref{e.mean-full}, \eqref{e.frechet-mean-vol-full}, or
\eqref{e.mean-vol}. In all cases, the average Lagrangian is given by
\eqref{e.L} and the expansion vector fields are expressed in terms of
fluctuations by \eqref{e.up-upp}.  In the following, we make a choice
that allows for further simultaneous treatment of the Euler equations
and Burgers' equations.  Below, we only assume that the fluctuation
vector fields satisfy the transport equation
\eqref{e.transport.a}. This is compatible with both definitions of the
map-average, equations \eqref{e.mean-full} and
\eqref{e.frechet-mean-vol-full}.  The case when the average map is
defined by \eqref{e.mean-vol} is considered in
Section~\ref{s.geodesic.vol}.

To simplify notation, we drop the $\beta$ indexes from now on, writing
e.g. $u'$ for $u_\beta'$ and $w$ for $w_\beta$, as no confusion can
result from such simplification.  Further, differentiating
\eqref{e.transport.a} in time, setting $\eps=0$, and substituting for
$\dot w$ from \eqref{e.first-order-taylor}, we can eliminate
$\dot w^{\prime}$ from \eqref{e.upp-reduced} to obtain
\begin{gather} \label{e.upp}
u'' 
= \nabla_{w} (\Lie_{u} w) \,  + \nabla_{\Lie_{u} w}  w \, 
- \Lie_{u} (\nabla_w w ) \,.
\end{gather} 
Regrouping terms and recalling the standard geometric identity
	\begin{gather} 
	\Lie_u w \equiv [u,w] = \nabla_u w - \nabla_w u \,,  
	\end{gather}
we further simplify \eqref{e.upp} as follows: 
\begin{equation} \label{e.upp1}
u''=-R(u,w)w+\nabla_{\nabla_w w} u - \nabla_w \nabla_w u \,.	
\end{equation}
Then, substituting \eqref{e.up} and \eqref{e.upp1} into \eqref{e.L},
we obtain
\begin{align}\label{e.Lavu}
  \bar L 
  & = \tfrac{1}{2}\int_\Omega |u|^2
      + \eps^2 \, g \bigl( \langle -R(u,w)w + \nabla_{\nabla_w w} u
        - \nabla_w \nabla_w u \rangle, u \bigr) \, \mu (x) \notag \\
  & \equiv L_0 + \tfrac{1}{2} \, \eps^2 \, L_2 \,.
\end{align}

\section{Isotropy of fluctuations}
\label{s.isotropy}

The final simplification of the averaged Lagrangian $L_2$ comes from
the isotropy assumption. Let $\{e_i=\partial/\partial {x_i} \}$ be a
set of coordinate vector fields and write
\begin{equation}
  w= w^{i} e_i \,.
\end{equation}
Statistical isotropy of fluctuations shall be expressed by the
condition
\begin{equation}\label{e.isotropy}
  \langle w^i w^j \rangle = g^{ij} \,,
\end{equation} 
where $g^{ij}$ are the components of the inverse metric tensor.  Under
this assumption, the terms in \eqref{e.Lavu} which contribute to the
$L_2$-Lagrangian simplify as follows.  First, using the Bianchi
identity, we compute
\begin{align}
  g(\langle R(u,w)w) \rangle,u) 
  & = \langle g( R(w,u)u ,w) \rangle\notag \\
  & = \langle g_{ij} R^i_{klm} w^k u^l u^m w^j \rangle \notag \\
  & = R^{i}_{ilm} u^l u^m \notag \\
  & = \Ric(u,u) \,.
  \label{e.term1}
\end{align}
Second, we find by direct computation that
\begin{align}
  \langle \nabla_w \nabla_w u - \nabla_{\nabla_w w} u \rangle
  & = \langle
        w^i \, \nabla_{e_i} (w_j \nabla_{e_j} u)
        - w^i \, \nabla_{\nabla_{e_i} (w^j e_j)} u
      \rangle
      \notag \\
  & = \langle
        w^i w^j \, \nabla_{e_i} \nabla_{e_j} u
        +  w^i \, (\nabla_{e_i} w_j) \, (\nabla_{e_j} u) 
        - w^i w^j \, \nabla_{\nabla_{e_i} e_j} u
        - w^i \, \nabla_{(\nabla_{e_i} w^j) e_j} u        
      \rangle
      \notag \\
  & = \langle w^i w^j \rangle \,
      ({\nabla_{e_i} \nabla_{e_j}} - \nabla_{\nabla_{e_i} e_j}) u 
      \notag \\
  & = \rolap u \,.
  \label{e.terms23}
\end{align}
Noting that the right hand sides of \eqref{e.term1} and
\eqref{e.terms23} in the metric inner product with $u$ add up to a
quadratic form involving the 
Ricci Laplacian, see \eqref{e.riclap},
we find that the averaged Lagrangian to second order in $\varepsilon$
reads
\begin{gather} \label{e.l-final}
 \bar  L = \tfrac12 \int_\Omega \lvert u \rvert^2 
      - \eps^2 \, g(\Delta_R u, u) \, \mu (x)\,.
\end{gather}
This is precisely the Lagrangian \eqref{e.LEuleraLap} of the EPDiff
and of the Euler-$\alpha$ equations.  The Camassa--Holm equations are
the EPDiff equations on a one-dimensional manifold.  For the latter,
it is easier, of course, to verify the passage from \eqref{e.Lavu} to
\eqref{e.l-final} directly in Euclidean coordinates.

\section{Intrinsic derivation of the Euler-$\alpha$ equations}
\label{s.geodesic.vol}

The derivation of Euler-$\alpha$ equations in
Sections~\ref{s.geodesic}--\ref{s.isotropy} uses the notion of mean
flow arising from connecting elements of $\SDiff (\Omega)$ by curves
lying in $\Ds (\Omega)$. A more natural definition would use the
notion of distance intrinsic to $\SDiff(\Omega)$. The argument below
shows that this intrinsic definition of the geodesic mean also leads
to the Euler-$\alpha$ equations.

From now on, we assume that $\eta$ is the Fr\'echet mean of $\eta_\fl$
in $\SDiff(\Omega)$ as specified by \eqref{e.mean-vol}, so that the
fluctuations satisfy the incompressible Euler equation
\eqref{e.transportvol}, see \cite{GilbertV:2016:GeometricGL}.  The
``pressure'' field $\phi_\eps$ is recovered by solving the Poisson
equation
\begin{subequations}
\begin{gather}
  \Delta \phi_\eps = -\Div (\nabla_{w_\eps} w_\eps )
  \quad \text{ in } \Omega \,,
\end{gather}
\end{subequations}
which we will write as $\phi_\eps= - \Delta^{-1} \Div (\nabla_{w_\eps}
w_\eps)$. 

Assuming the Taylor hypothesis \eqref{e.first-order-taylor} and the
isotropy of fluctuations \eqref{e.isotropy}, the calculation from
Sections~\ref{s.geodesic}--\ref{s.isotropy} are modified as follows.
Fluctuations now satisfy the Euler equations \eqref{e.transportvol.a}
rather than the transport equation \eqref{e.transport.a} so that
\eqref{e.upp} is replaced by
\begin{gather} \label{e.uppvol}
  u'' 
  = \nabla_{w} (\Lie_{u} w) + \nabla_{\Lie_{u} w}  w 
    - \Lie_{u} (\nabla_w w ) - \Lie_{u} \phi - \nabla \dot \phi \,.
\end{gather} 
Therefore, the expression of the $L^2$-Lagrangian derived in
Section~\ref{s.isotropy} must be augmented with two extra terms, so
that
\begin{align}
  L_2
  & = - \biggl\langle
          \int_\Omega g \bigl( \Delta_R u + \Lie_u \nabla \phi
            + \nabla \dot \phi ,u \bigr) \, \mu (x)
        \biggr\rangle
      \notag \\
  & = - \int_\Omega g (\Delta_R u, u) \,  \mu (x)
      - \biggl\langle
          \int_{\Omega } g(\Lie_u \nabla \phi, u ) \, \mu (x)
        \biggr\rangle \,,
  \label{e.l2vol2}
\end{align}
where the last term in the first line vanishes since gradients are
$L^2$-orthogonal to divergence free vector fields.

We compute the last term in \eqref{e.l2vol2} by noting that due to the
Hodge decomposition, the operator $\nabla \Delta^{-1} \Div$ is $L^2$
symmetric, i.e., for arbitrary sufficiently smooth vector fields $v$
and $w$,
\begin{equation}
  \int_{\Omega} g(\nabla \Delta^{-1} \Div (v), w ) \, \mu (x)
  = \int_{\Omega} g(v, \nabla \Delta^{-1} \Div (w)) \, \mu (x) \,.
\end{equation}
Since $u$ is necessarily divergence free as a vector field generating $\eta \in \SDiff(\Omega)$, integrating by parts, we have  
\begin{align} \label{e.lugphi}
  \int_{\Omega} g(\Lie_u \nabla \phi,u) \, \mu (x)
  & = \int_\Omega g(\nabla_u \nabla \phi,u) -
        g(\nabla_{\nabla \phi} u,u) \, \mu (x) \notag \\
  & = -\int_\Omega g\bigl(\nabla \phi \,, \nabla_u u
        + \tfrac{1}{2} \nabla |u|^2 \bigr) \, \mu (x) \notag \\
  & = \int_\Omega g\bigl( \nabla_w w, \nabla \Delta^{-1} \Div
        (\nabla_u u+\tfrac{1}{2} \nabla |u|^2) \bigr) \, \mu (x) \notag \\
  & = -\int_\Omega g\bigl( w, \nabla_w \nabla \Delta^{-1} \Div
        (\nabla_u u+\tfrac{1}{2} \nabla |u|^2) \bigr) \, \mu (x) \,.
\end{align}
For an arbitrary vector field $v$, 
\begin{align}\label{e.lapphi}
  \biggl\langle
    \int_{\Omega} g(w, \nabla_w v) \, \mu (x)
  \biggr\rangle 
  & = \int_{\Omega} g_{ij} \, \langle w^i w^k \rangle \,
      \biggl(
        \frac{\partial v^j}{\partial x_k} + \Gamma^j_{ks}v^s
      \biggr) \, \mu (x) \notag \\
  & = \int_\Omega
      \biggl(
        \frac{\partial v^j}{\partial x_j} + \Gamma^j_{js}v^s
      \biggr) \, \mu (x) \notag \\
  & = \int_\Omega \Div v \, \mu (x) \,,
\end{align}
where the last equality follows from the standard expression for the
divergence of a vector field,
\begin{equation}
  \Div v = \frac{1}{\sqrt{g}} \, \frac{\partial}{\partial x_i}
  \bigl( \sqrt{g} \, v^i \bigr) \,.
\end{equation}
Now, combining \eqref{e.lugphi} and \eqref{e.lapphi}, we obtain
\begin{align}
  \label{e.l2-4}
  \biggl\langle
    \int_{\Omega} g(\Lie_u \nabla \phi,u)  \, \mu (x)
  \biggr\rangle
  = -\int_{\Omega} \Div( \nabla_u u+\tfrac{1}{2} \nabla |u|^2)  \, \mu (x)
  =0 \,.
\end{align}
Substituting \eqref{e.l2-4} into \eqref{e.l2vol2}, we obtain 
\begin{equation}
L_2=-\int_{\Omega} g( \Delta_R u,u) \, \mu (x)\,,
\end{equation}
so that the full averaged Lagrangian $\bar L$ coincides with the
Euler-$\alpha$ Lagrangian \eqref{e.LEuleraLap}.

\section{Averaged equations of motion}
\label{s.ep}

In this section, we derive Euler-$\alpha$ equations \eqref{e.eulera}
and the EPDiff equations \eqref{e.epdiff} as the Euler--Poincar\'e
equations for the averaged Lagrangian $\bar L$ on $\SDiff(\Omega)$ and
$\Ds(\Omega)$, respectively.  To do so, we must compute the stationary
points of the averaged action
\begin{gather} 
\bar S = \int_{t_1}^{t_2} \bar L(\dot \eta, \eta) \, \d t
\label{e.avaction}
\end{gather}
with respect to variations of the flow map $\delta \eta$ in the
respective configuration spaces which vanish at the temporal endpoints.

First, we note that variations in the flow map
$\delta \eta = w \circ \eta$ and the fluid velocity
$u = \dot \eta \circ \eta^{-1}$ are related by the Lin constraint
\citep{Bretherton:1970:NoteHP}
\begin{equation} \label{e.lin}
	\delta u = \dot w + \Lie_u w \,,
\end{equation}
which is proved analogously to \eqref{e.uprime.a}.  Next, due to the
symmetry of the Ricci tensor, the averaged Lagrangian $\bar L$ is of
the form
\begin{equation}
\bar L = \tfrac{1}{2}\int_\Omega g(Au,u) \, \mu(x) \,,
\end{equation}
where 
\begin{equation}
A=\operatorname{id}-\eps^2 \Delta_R 
\end{equation}
is a linear $L^2(\Omega,g)$-self-adjoint operator on vector
fields. Therefore,
\begin{align} \label{e.varl}
	\delta L &= \int_\Omega g(Au,\delta u) \, \mu(x) 
	=\int_\Omega g(m,\delta u) \, \mu(x) \,,
\end{align} 
where the circulation velocity $m$ is given by 
\begin{equation}
m=Au = u-\eps^2 \Delta_R u \,.
\end{equation}
From this point on, while the overall strategy remains similar, the
details of computation depend on the configuration space. Therefore,
we will treat both cases separately.
  
On $\Ds(\Omega)$, $w$ is an arbitrary vector field in $V$.  Using
\eqref{e.lin}, \eqref{e.varl}, and integration by parts, we compute
\begin{align}\label {e.varS}
  \delta \bar S
  & = \int_{t_1}^{t_2} \delta \bar L \, \d t
    = \int_{t_1}^{t_2} \int_\Omega g(m,\delta u) \, \mu(x) \, \d t
      \notag \\
  & = \int_{t_1}^{t_2}
      \int_\Omega g(m,\dot w + \Lie_u w) \, \mu(x) \, \d t \notag \\
  & = \int_{t_1}^{t_2} \int_\Omega -g(\dot m,w) +g(m,\nabla_u w)
      - g(m,\nabla_w u) \, \mu(x) \, \d t \notag \\
  & = - \int_{t_1}^{t_2} \int_\Omega g(\dot m,w)+g(\nabla_u m, w)
      - \nabla_u g(m, w)+ g((\nabla u)^T m,w ) \, \mu(x) \, \d t \notag \\
  & = -\int_{t_1}^{t_2} \int_\Omega g(\dot m+\nabla_u m
      + m\Div u+(\nabla u)^T m,w) \, \mu(x) \, \d t
    \equiv 0 \,.  
\end{align}
Since $w$ is arbitrary, $m$ must satisfy the EPDiff momentum equation \eqref{e.epdiffa}. 

On $\SDiff(\Omega)$, $w$ is an arbitrary divergence-free vector field
in $\vfd$.  Moreover, the velocity $u$ is a curve in $\vfd$.
Therefore, the computation in \eqref{e.varS} implies that
\begin{equation} 
  \int_{t_1}^{t_2} \int_\Omega g(\dot m+\nabla_u m
  + (\nabla u)^T m,w) \, \mu(x) \, \d t =0 \,.
\end{equation}
Since, by the Hodge decomposition, the space of vector fields
orthogonal to $\vfd$ in $L^2(\Omega,g)$ consists of gradients, the
circulation velocity $m$ satisfies the Euler-$\alpha$ momentum
equation \eqref{e.euleraa}.

\section{Manifolds with boundaries}
\label{s.boundaries}

Our methods are flexible enough to treat manifolds with a
boundary. However, it has already been noted in
\cite{MarsdenS:2003:AnisotropicLA} that natural for Euler equations
no-flux boundary conditions $u \cdot n =0$, where $n$ stands for the
outward normal to the boundary $\partial \Omega$, are incompatible
with isotropy of fluctuations. 

Indeed, suppose that the normal $n(x)=e_3$ at a point
$x \in \partial \Omega$.  Then, the no-flux boundary conditions would
imply $w_3(x)=0$ for an arbitrary fluctuation vector field $w$, so
that $\langle w_3 (x) \otimes w_3 (x) \rangle=0$, whereas isotropy
requires $\langle w_3 (x) \otimes w_3 (x) \rangle=1$.  A similar
problem emerges for Burgers' equations with the natural no-slip
boundary condition $u=0$ on $\partial \Omega$.

Thus, on manifolds with boundary, one must generally consider
anisotropic equations, which are a coupled system of evolution
equations for the mean velocity and Taylor diffusivity tensor
\begin{equation}
\kappa = w \otimes w \,,
\end{equation}
see, e.g., \cite{MarsdenS:2003:AnisotropicLA} or
\cite{holm1999fluctuation}.

However, for certain simplified geometries, for instance for a
horizontal strip $\Omega=\R^2 \times [0,H]$, the rigid lid boundary
conditions are compatible with spatial uniformity of the Taylor
diffusivity tensor $\kappa$. In such cases, one could still derive
analogues of isotropic Euler-$\alpha$ equations on manifolds with
boundary by replacing the isotropy with an appropriate form of spatial
uniformity in the closure hypothesis. We refer the reader to
\cite{BadinOV:2017:LagrangianAE} for the examples of such a
construction.

\section*{Acknowledgments}

We thank Gualtiero Badin and Jacques Vanneste for many interesting
discussions.  This paper contributes to the project ``The Interior
Energy Pathway'' of the Collaborative Research Center TRR 181 ``Energy
Transfers in Atmosphere and Ocean'' funded by the German Research
Foundation.  Funding through the TRR 181 is gratefully acknowledged.

\bibliographystyle{apacite}
\renewenvironment{APACrefURL}[1][]{}{}
\def\url#1{}

\end{document}